\documentclass[twocolumn]{aastex61}
\usepackage{mathrsfs, mathtools}
\newcommand{\as}{\arcsec{}}
\usepackage{blindtext}

\begin{document}

\title{An Inner Disk in the Large Gap of the Transition Disk SR\,24S}
\correspondingauthor{Paola~Pinilla, Hubble Fellow}
\email{pinilla@email.arizona.edu}

\author{Paola Pinilla}
\affiliation{Department of Astronomy/Steward Observatory, The University of Arizona, 933 North Cherry Avenue, Tucson, AZ 85721, USA}
\affiliation{Hubble Fellow}

\author{Myriam Benisty}
\affiliation{Unidad Mixta Internacional Franco-Chilena de Astronom\'{i}a (CNRS, UMI 3386), Departamento de Astronom\'{i}a, Universidad de Chile, Camino El Observatorio 1515, Las Condes, Santiago, Chile}
\affiliation{Univ. Grenoble Alpes, CNRS, IPAG, 38000 Grenoble, France}

\author{Paolo Cazzoletti}
\affiliation{Max-Planck-Institute for Extraterrestrial Physics (MPE), Giessen- bachstr. 1, 85748, Garching, Germany}

\author{Daniel Harsono}
\affiliation{Leiden Observatory, Leiden University, P.O. Box 9513, 2300 RA Leiden, The Netherlands}

\author{Laura M.~P\'erez}
\affiliation{Departamento de Astronom\'ia, Universidad de Chile, Camino El Observatorio 1515, Las Condes, Santiago, Chile}

\author{Marco Tazzari}
\affiliation{Institute of Astronomy, University of Cambridge, Madingley Road, Cambridge CB3 0HA, UK}

\begin{abstract}
We report new Atacama Large Millimeter/sub-millimeter Array (ALMA) Band 3 observations at 2.75\,mm of the TD around SR\,24S with an angular resolution of $\sim$0.11\as$\times$ 0.09\as and a peak signal-to-noise ratio of $\sim24$. We detect an inner disk and a mostly symmetric ring-like structure that peaks at $\sim$0.32\as, that is $\sim$37\,au at a distance of $\sim$114.4\,pc.  The full width at half maximum of this ring is $\sim$28\,au. We analyze the observed structures by fitting the dust continuum visibilities using different models for the intensity profile, and compare with previous ALMA observations of the same disk at 0.45\,mm and 1.30\,mm. We qualitatively compare the results of these fits with theoretical predictions of different scenarios for the formation of a cavity or large gap. The comparison of the dust continuum structure between different ALMA bands indicates that photoevaporation and dead zone can be excluded as leading mechanisms for the cavity formation in SR\,24S disk, leaving the planet scenario (single or multiple planets) as the most plausible mechanism. We compared the 2.75\,mm emission with published (sub-)centimeter data and find that the inner disk is likely tracing dust thermal emission. This implies that any companion in the system should allow dust to move inwards throughout the gap and replenish the inner disk. In the case of one single planet, this puts strong constraints on the mass of the potential planet inside the cavity and the disk viscosity of about $\lesssim$5\,$M_{\rm{Jup}}$ and $\alpha\sim10^{-4}-10^{-3}$, respectively.
\end{abstract}


\section{Introduction}     \label{introduction}

Recent high-angular resolution observations, with for example the Atacama Large Millimeter/sub-millimeter Array (ALMA), have revolutionized the field of planet formation by unveiling a variety of structures observed in protoplanetary disks. In general, when observing at high angular resolution, ALMA has identified two broad categories of disks. To the first category belong those disks with multiple rings and gaps \citep[e.g.,][]{andrews2018, long2018}, while in the second category are those with a large dust gap or cavity \citep[e.g.,][]{marel2018, pinilla2018}. The second category are transition disks (TDs), where the observed cavities are usually surrounded by a ring-like structure that may or not be axisymmetric. Interestingly, some of these disks appear to also have more complex structures in the dust, beyond a simple cavity and ring structure \citep[e.g.,][]{dong2018}. It is therefore possible that, in the near future,  the distinction between the two categories will become less evident.

Nevertheless, TDs were already identified three decades ago, prior to spatially resolved any kind of substructures. They were recognized by their spectral energy distributions (SEDs), which show weak near- and mid-infrared excess emissions, but substantial excess beyond 20 microns \citep{strom1989}. This type of SED suggested the presence of dust-depleted cavities, which were later spatially resolved at different wavelengths \citep[e.g.][]{brown2009, andrews2011, espaillat2014}. Therefore, TDs have been for years excellent laboratories to investigate disk evolution. The study of such objects is important to make a step forward in the current understanding of the origin of more complex structures that we observe today. 

The ring-like shape of TDs may result from the trapping of dust particles in specific regions of protoplanetary disk known as pressure bumps. These pressure bumps were already proposed by \cite{wipple1972} to overcome one of the most challenging problems of planet formation: the radial drift barrier \citep{weidenschilling1977}. In a protoplanetary disk with a homogeneous gas distribution, dust particles feel an aerodynamic drag that causes them to migrate inward, in particular when they are millimeter and centimeter in size, and they are in the outer parts of the disk (beyond $\sim$20\,au). This radial drift has been a challenge for understanding observations of protoplanetary disks, which show that millimeter-sized particles remain in the outer disk for millions of years, despite radial drift \citep[see e.g.,][]{ricci2010b, testi2014}. Pressure bumps provide a solution to the drift barrier because the aerodynamical drag between the dust particles and the gas is reduced or totally suppressed near or at the pressure maximum. 

Several observational tests can be performed to inspect if particle trapping is occurring in protoplanetary disks. For example, \cite{dullemond2018b} investigated if dust trapping is operational in disks with substructures observed at high angular resolution with ALMA \citep{andrews2018}, by comparing the width of the dust rings with the width of a potential gas pressure bump. If the width of the dust ring is lower than the width of the pressure bump, it is likely that trapping is in action. Measuring the width of a gas pressure bump directly from observations is still challenging, although it has been possible in very few cases \citep[e.g.,][for the case of HD\,163296]{teague2018}.

One consequence of dust trapping is that dust particles that are more decoupled from the gas (but not totally decoupled) are subject to feel the radial drift more efficiently. As a result,  in a pressure maximum centimeter particles are more concentrated than millimeter particles, which in turn will be more concentrated than micron-sized particles. This prediction can be tested by observing disks at different wavelengths that trace different grain sizes. Depending on the origin of the pressure bumps, different degrees of segregation in the distribution of small, intermediate and large particles are expected \citep[see e.g.,][for a review]{pinilla_youdin2017}. For instance, a massive planet has been invoked by theorists to explain the large cavities in TDs \citep[e.g.,][]{rice2006, paardekooper2006}. In this case, one direct consequence of dust trapping is that the observed dust continuum cavity increases in size with increasing wavelength \citep[e.g.,][]{ovelar2013}. The majority of TDs that have been observed at near-infrared scattered light and millimeter wavelengths have shown this kind of segregation \citep{villenave2019}, hinting to embedded planets in the cavities. Observational campaigns have been carried out to search for companions in TDs \citep[e.g.,][]{cugno2019}, and few objects have planet candidates, although with several controversies whether or not the observed emission comes from a point source or from the disk itself \citep[e.g.,][]{quanz2013, rameau217, reggiani2018}. So far, only one companion in a TD has been confirmed \citep[PDS\,70b,][]{keppler2018}.

A few of the other proposed mechanisms to open a large gap or cavity in disks are photoevaporation \citep[e.g.,][]{alexander2007, ercolano2017} and the presence of regions of low ionization where the magnetorotational instability is inhibited \citep[the so called dead-zone, e.g.,][]{flock2015}. The dust segregation is expected to vary (and hence the cavity size at different wavelengths) depending on the origin of the pressure bump. It is therefore fundamental to obtain multiwavelength observations of disks with substructures, including TDs, in order to understand the physical mechanisms that are allowing the millimeter-sized particles to persist in the outer disk, as presented in this paper.

In this paper, we present new ALMA Band 3 observations at 2.75\,mm of the TD around SR\,24S and compare our data with published ALMA observations of the same disk at 0.45\,mm and 1.30\,mm. We use this multiwavelength comparison to better understand the potential origin of the cavity observed in SR\,24S. The paper is organized as follows. In Sect.~\ref{sect:observations}, we describe the details of our ALMA observations. The results and analysis of our data, together with the comparison with previous ALMA observations are presented in Sect.~\ref{sect:results_analysis}. The discussion and conclusions are in Sect.~\ref{sect:discussion} and Sect.~\ref{sect:conclusions}, respectively.

\section{Observations} \label{sect:observations}

\begin{figure*}
 \centering
 \setlength{\tabcolsep}{0.1pt}
  \begin{tabular}{cc}   
   	\includegraphics[width=9.0cm]{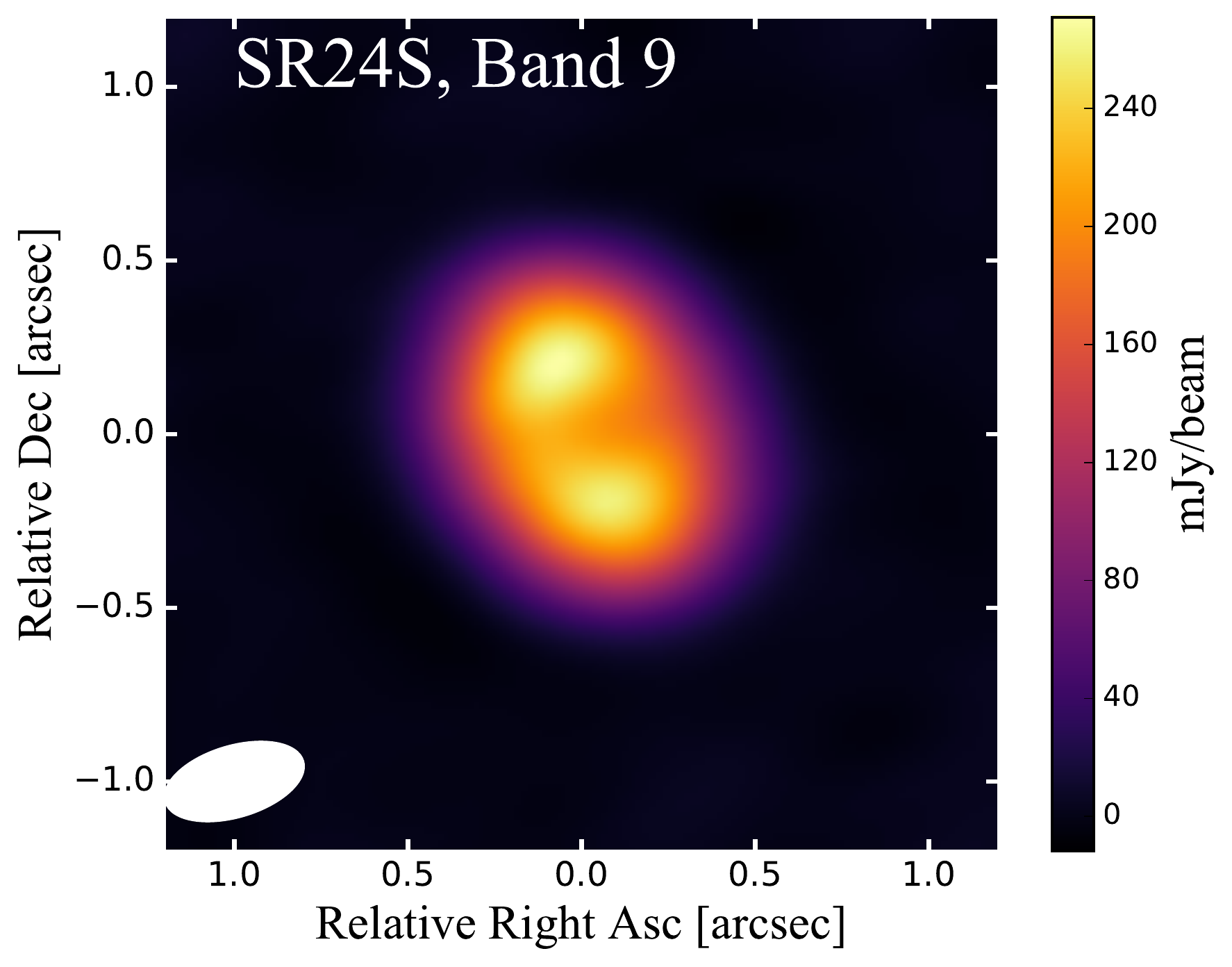}&
	\includegraphics[width=9.0cm]{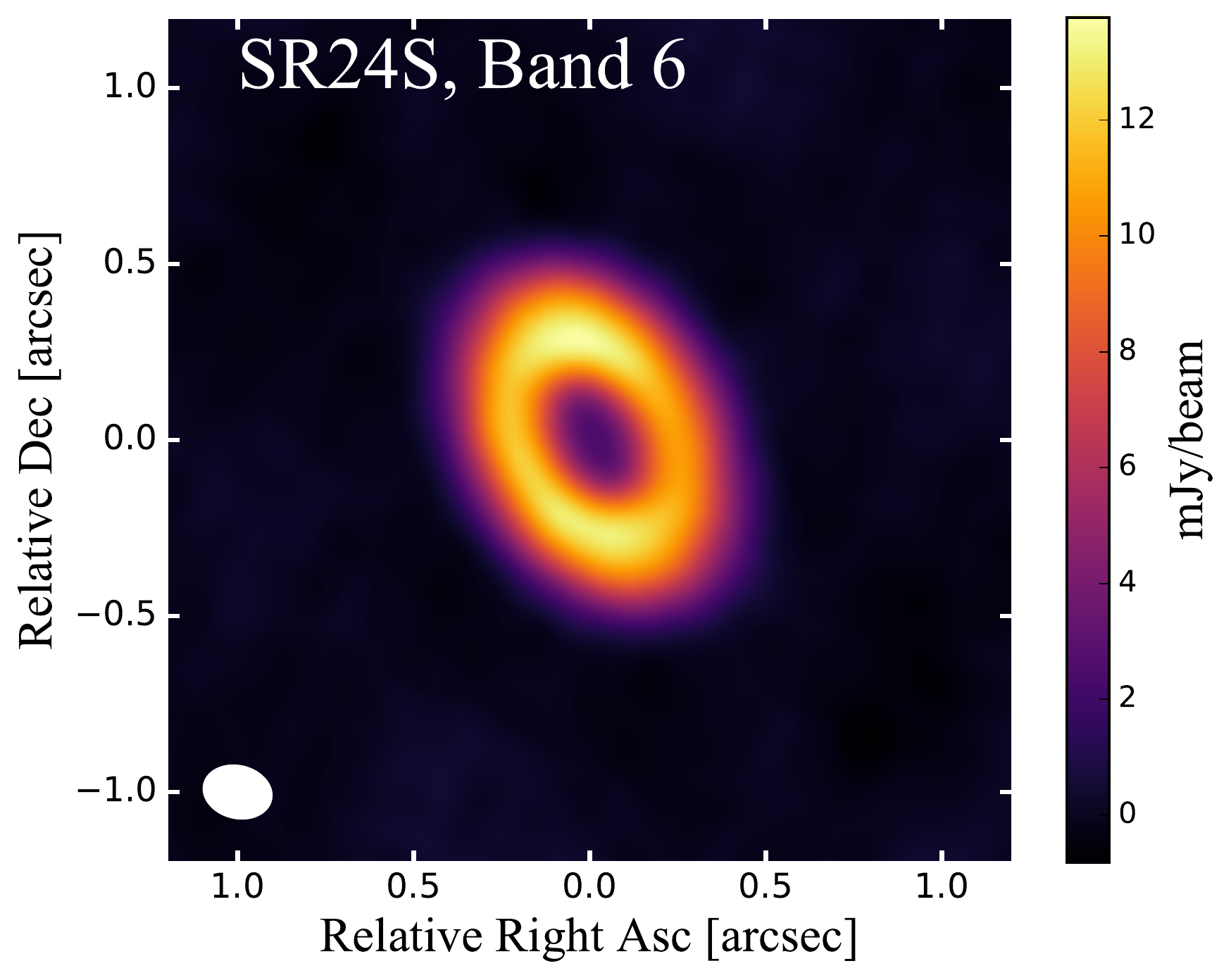}\\
	\includegraphics[width=9.0cm]{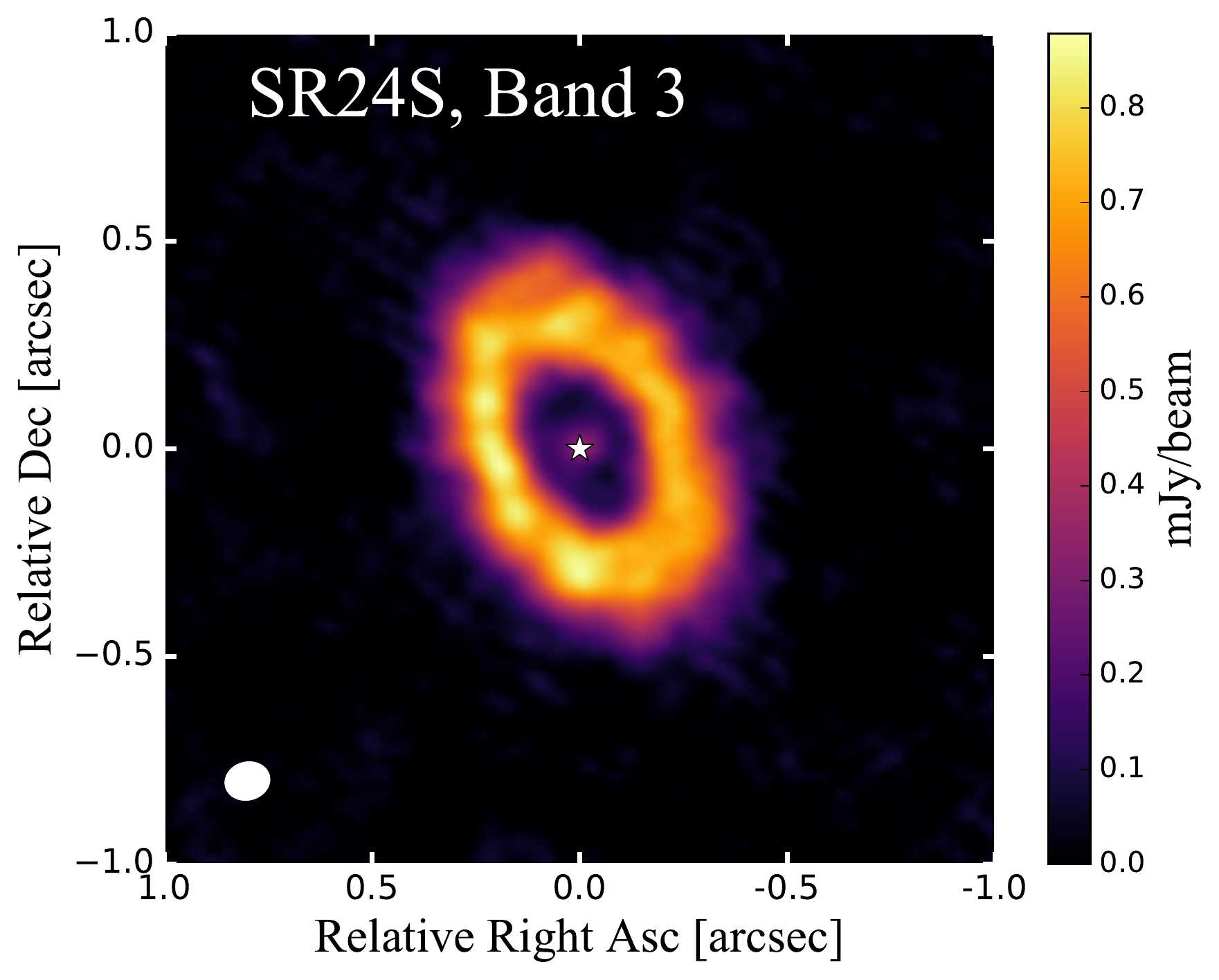}&
	\includegraphics[width=9.0cm]{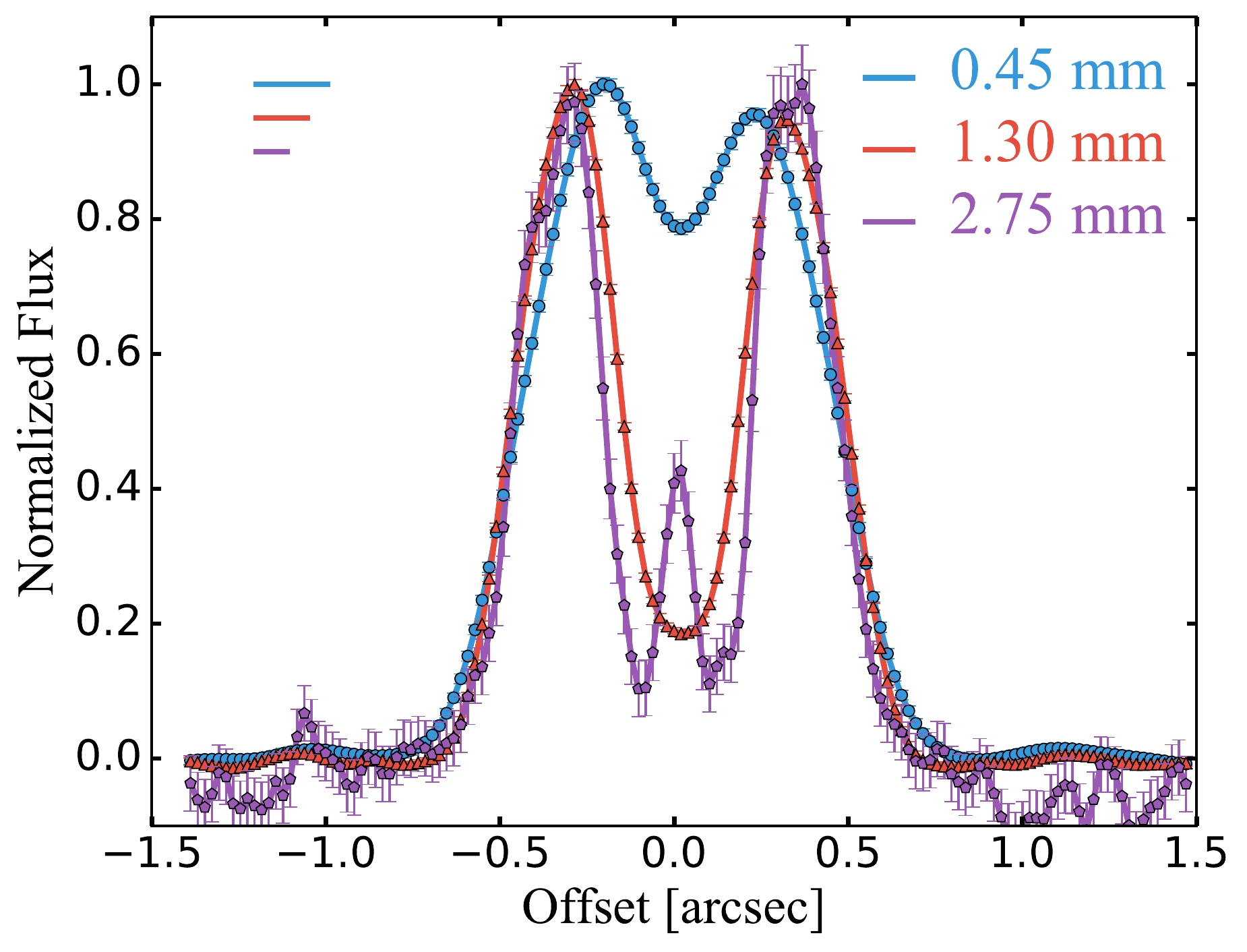}
   \end{tabular}
   \caption{ALMA observations of the disk around SR\,24S in Band 9 or 0.45\,mm (upper left panel), in Band 6 or 1.3\,mm (upper right panel), and in Band 3 or 2.75\,mm (bottom left panel). In each panel, the beam is shown in the bottom left of the image. The bottom right panel corresponds to the intensity profile as a function of offset at the disk PA ($\sim27^{\circ}$), errors correspond to the rms of each observation. The horizontal bars represent the minor axis of the synthesized beams.}
   \label{ALMA_SR24S}
\end{figure*}

SR\,24S was observed with ALMA in Band 3 during Cycle 5 on November 9th, 2017 (\#2017.1.00884.S).  For these observations 43 antennas were used, with a baseline range from 138\,m to 13894.4\,m. The source was observed in four spectral windows, two of them centered at $107.9$\,GHz, one at $110.2$\,GHz, and the other one at $109.8$\,GHz; for a mean frequency of $\sim$109\,GHz or a wavelength of 2.75\,mm. The quasar QSO J1427-4206 was observed for bandpass and flux calibration, while the quasar QSO J1625-2527 was observed for phase calibration. The total observing time was 22.5\,mins, with a total on-source time of $\sim$8\,min. We also aimed to detect $^{13}$CO(1-0) at 110.20\,GHz and C$^{18}$O (1-0) at 109.78\,GHz, but did not get a clear detection of these emission lines. We performed self-calibration, which slightly improved the signal-to-noise ratio of the data compared to the delivered data. The data were calibrated using the Common Astronomy Software Package \citep[CASA,][]{mcmullin2007}.

Before imaging, the data were correctly centered by fitting a simple Gaussian to the data, using {\tt uvmodelfit}. The obtained center was $\alpha_{2000}$=16:26:58.51, $\delta_{2000}$=-24:45:37.24, which was used to correct the phase center and obtain the visibilities using {\tt fixvis}. From the fitting,  the position angle (PA) and inclination were $26.8^{\circ}\pm{1.3}^{\circ}$ and $47.6^{\circ}\pm{2.4}^{\circ}$ respectively, in agreement with previous observations \citep{marel2015, fernandez2017, pinilla2017}.

Continuum imaging was performed using the {\tt clean}  algorithm. We used the natural weighting scheme to obtain the best sensitivity possible. The final beam size was 0.106\as$\times$0.088\as, achieving a rms of $\sim$38\,$\mu$Jy\,beam$^{-1}$. The total flux density and the peak brightness from the image is 28.9\,mJy and 0.9\,mJy\,beam$^{-1}$, respectively. This implies a signal-to-noise ratio with respect to the peak of $\sim24$. 

The details of the ALMA Cycle 0 and Cycle 2 observations (0.45\,mm and 1.30\,mm, respectively) and the respective calibrations are explained in \cite{marel2015} and \cite{pinilla2017}.

\section{Results and Data Analysis} \label{sect:results_analysis}

\subsection{Dust morphology and comparison with previous ALMA observations}

\begin{figure*}
 \centering
 \setlength{\tabcolsep}{0.1pt}
  \begin{tabular}{cc}   
   	\includegraphics[width=9.0cm]{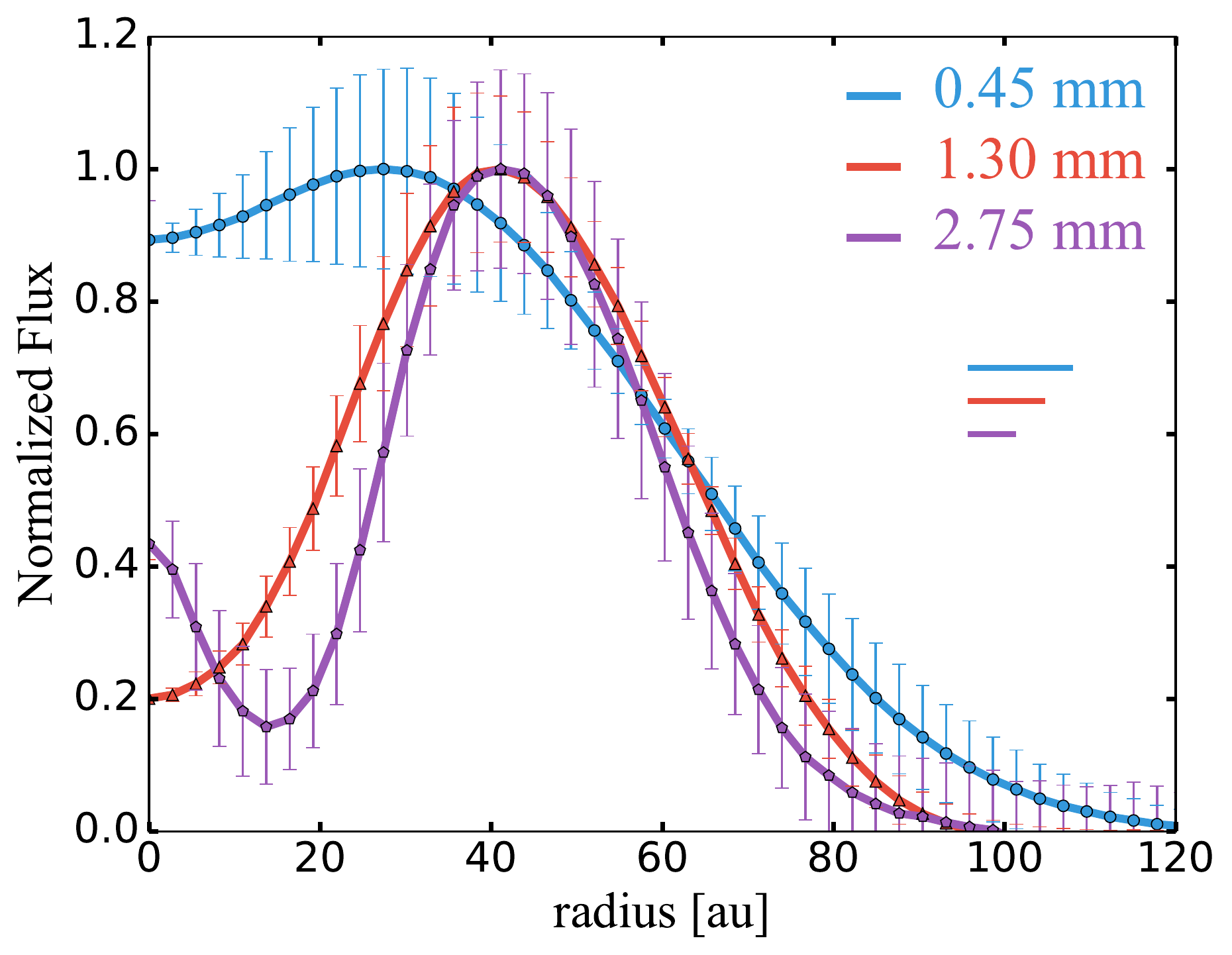}&
	\includegraphics[width=9.0cm]{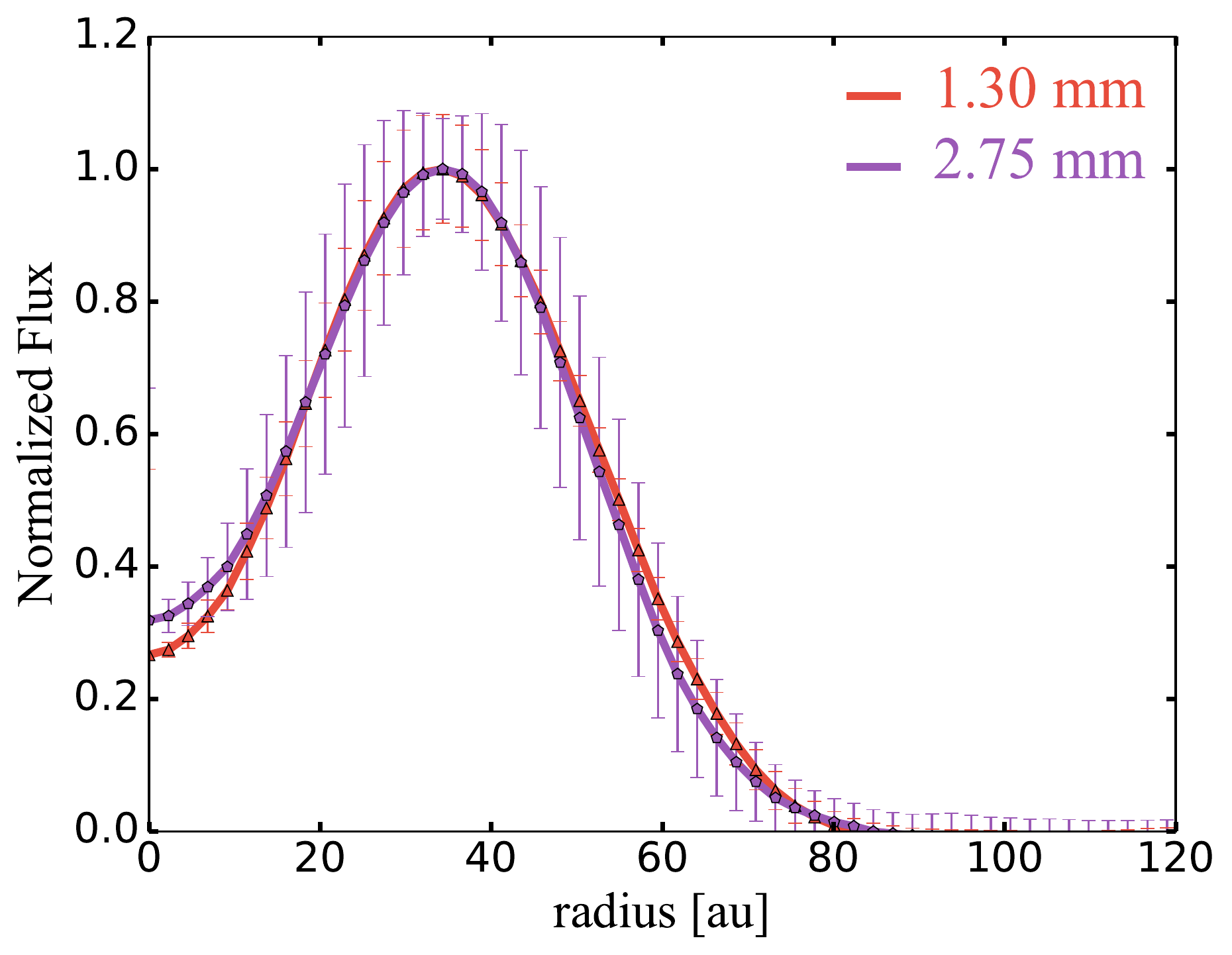}
   \end{tabular}
   \caption{Left: Radial profiles after azimuthally averaging the deprojected images (assuming d=114.4 pc, errors include standard deviation in each radial bin and the rms from the observations). The horizontal bars represent half of the minor axis of the synthesized beams. Right: As left, but the images at 1.3\,mm and 2.75\,mm have been produced with the same circular beam of 0.19\as for comparison.}
   \label{radial_profiles}
\end{figure*}

The final image from our Band 3 observations is shown in the bottom left panel of Fig.~\ref{ALMA_SR24S}. In this figure, we include for comparison the previous ALMA observations (using the same cleaning procedure) at 0.45\,mm and 1.3\,mm obtained in Cycle 0 and Cycle 2, with a resolution of 0.37\as$\times$0.19\as and 0.19\as$\times$0.15\as, respectively. In addition, this figure shows the radial cut of the continuum flux, along the disk PA, and normalized to the peak, for each wavelength. 

A dust depleted cavity is resolved at the three wavelengths. However, the width of the ring is only spatially resolved in the Band 6 and in the Band 3 observations, as discussed in Sect.~\ref{ring_fit}. To check if the emission inside the ring is optically thin or thick, we calculate the optical depth as $\tau=-\ln[1-T_{\rm{brightness}}/T_{\rm{physical}}]$, with $T_{\rm{brightness}}$ and $T_{\rm{physical}}$ being the brightness and physical temperature respectively. The brightness temperature is calculated from the blackbody Planck function without assuming the Rayleigh-Jeans regime. The optical depth is higher than unity inside the ring for the Band 9 and for the Band 6 observations, as demonstrated in \cite{pinilla2017} (their Figure 7). While for the Band 3 observations, the emission inside the ring is optically thin, with a maximum value of $\tau$ at the peak of emission of 0.37 (when assuming a physical temperature of 20\,K).

The data in Band 3 shows emission from the inner disk. The total flux density inside a circular aperture of 75\,mas radius is $\sim$0.46\,mJy and the maximum is $\sim$0.33\,mJy\,beam$^{-1}$, that is a detection of about $\sim$7\,$\sigma$, which is clearly seen in the radial cut along the PA of the disk. In these observations, the inner disk seems to be centered around the central star, and we assumed in Sect.~\ref{ring_fit} a central Gaussian or a point source at the center to fit the emission of this inner disk.

The left panel of Fig.~\ref{radial_profiles} shows the azimuthally averaged radial profiles from the deprojected images and assuming a distance of 114.4\,pc $\pm$4.8\,pc \citep{gaia2018}. The errors include the standard deviation in each radial bin and the rms from the observations. This figure shows that the ring of emission at 0.45\,mm peaks around 30\,au, while it peaks at around 40\,au at 1.3\,mm and 2.75\,mm.  Since the resolution of the Band 6 and Band 3 images is similar (0.19\as$\times$0.15\as vs. 0.106\as$\times$0.088\as, respectively), we imaged with the same circular beam of 0.19\as and deprojected the two data sets  for comparison (right panel of Fig.~\ref{radial_profiles}). At this resolution, the emission coming from the inner disk in Band\,3 is not detectable in the image because of beam dilution, and the shape of the ring of emission is very similar to that in Band 6.

\subsection{Disk dust mass and spectral index} \label{dust_spectral}

Assuming optically thin emission, the dust disk mass can be calculated as $M_{\mathrm{dust}}\simeq\frac{{d^2 F_\nu}}{\kappa_\nu B_\nu (T(r))}$ \citep{hildebrand1983}. Considering a distance to the source of 114.4\,pc $\pm$4.8\,pc, a mass absorption coefficient at a given frequency given by $\kappa_\nu=2.3\,$cm$^{2}$\,g$^{-1}\times(\nu/230\,\rm{GHz})^{0.4}$ \citep[][]{andrews2013}, and a dust temperature of 20\,K \citep[e.g.][]{pascucci2016}; the total dust mass obtained from the total flux at 2.75\,mm is 55.3\,$M_{\oplus}\pm$7.3\,$M_{\oplus}$, when taking the total flux from the image. This uncertainty includes the uncertainty on the distance and 10\% of uncertainty from absolute flux calibration in addition to the rms of the data. The potentially large uncertainty in the dust opacity is not taken into account. This dust mass is lower than  the values obtained from ALMA observations at shorter wavelengths: using the new Gaia distance, the calculated dust mass using the 1.3\,mm observations is  85.8\,$M_{\oplus}\pm$14.1\,$M_{\oplus}$. The difference in dust mass might arise from spatial-filtering of the extended flux due to the lack of short baselines. In our observations, the maximum recoverable scale (MRS) is 1.1\as. As part of the same program, the TD around HD\,135344B was observed with similar resolution and MRS \citep{cazzoletti2018}. For HD\,135344B (which is more radially extended), we obtained short baselines observations to recover the large scales. The total flux of this source after combining short and long baselines is similar than when assuming only the long baselines observations. It is unclear if this would be the case for SR\,24S. However,
the 2.75\,mm flux from our observations is in good agreement with the flux obtained from the Australia Telescope Compact Array (ATCA) observations at 3\,mm of 26.6\,mJy \citep{ricci2010}, which synthesized beam has a full width at half maximum of $\sim$3-7\as.
Another possibility for the difference of $M_{\mathrm{dust}}$ is the opacity, which may have a more complex dependence on grain size and wavelength \citep{birnstiel2018} than what we assumed.

Using the total flux at 1.3\,mm of 220\,mJy \citep{pinilla2017}, and the 2.75\,mm flux, we find that the spatially integrated spectral index is 2.7$\pm$0.03 (error includes 10\% from flux calibration in addition to the rms of the observations). This value of the spectral index ($\alpha_{\mathrm{mm}}$) is in agreement with previous work \citep[e.g.,][]{zapata2017}, and it is higher than the value reported in \cite{pinilla2017} using the 0.45\,mm observations, which is dominated by optically thick dust emission. \cite{pinilla2014} found a positive correlation between the spatially integrated spectral index and the cavity size of TDs. Using such correlation ($\alpha_{\mathrm{mm}}=a\times r_{\mathrm{cav}}+b$, with $a=0.011\pm{0.007}$ and $b=2.36\pm{0.28}$), the expected cavity size is 31.0\,au$\pm$1.4\,au. This value is in agreement (within the resolution of the data) with the cavity size resolved in ALMA observations (Table~\ref{table:MCMC_fit}), as discussed below.   

\subsection{Fit of the dust morphology in the visibility plane} \label{ring_fit}

\begin{figure*}
 \centering
   	\includegraphics[width=18cm]{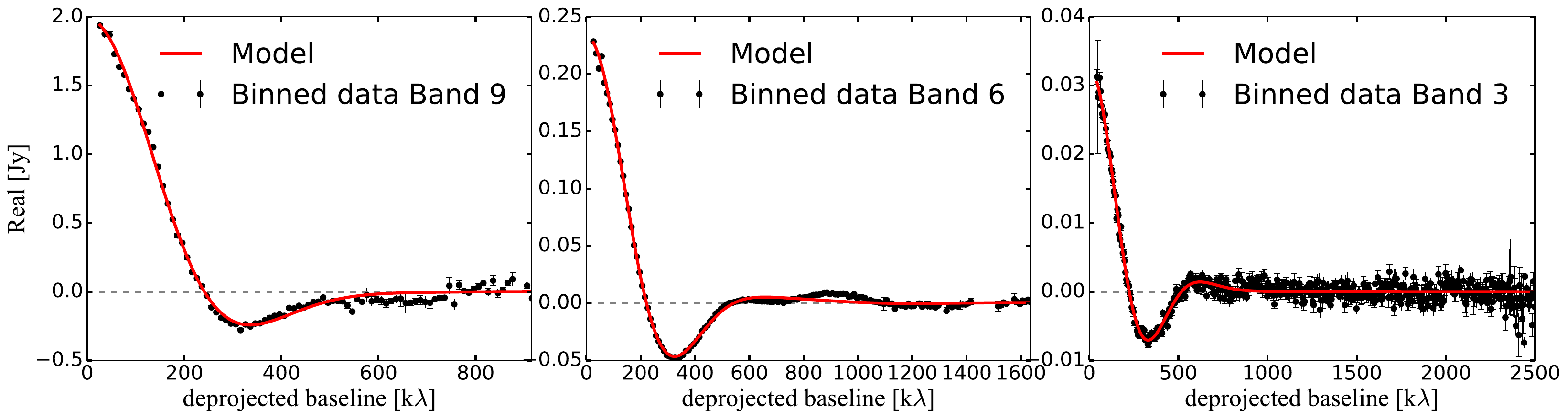}\\
   	\includegraphics[width=18cm]{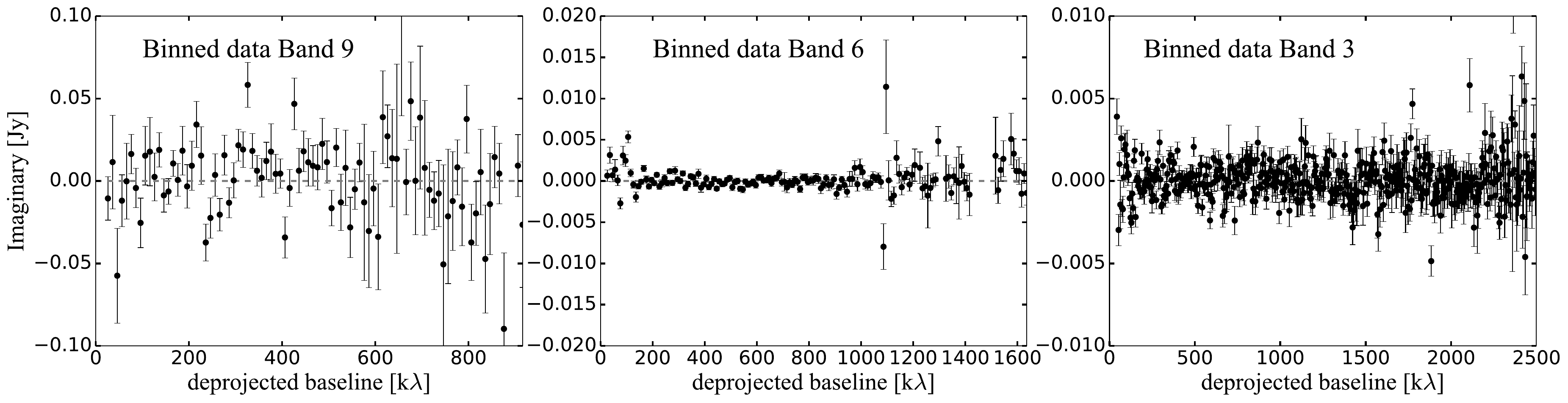}
   \caption{Top panels: Best model fit (model with the lowest BIC, see Table~\ref{table:MCMC_fit}) vs. the real part of the binned and deprojected visibilities for Band 9 (0.45\,mm, left panel), Band 6 (1.3\,mm, middle panel), and Band 3 (2.75\,mm, right panel). The error bars correspond to the standard error in each bin. Bottom panels: Imaginary part of the visibilities for each band after centering the target.}
   \label{best_fit_models}
\end{figure*}

To fit the millimeter dust continuum emission at the three wavelengths, we perform an analysis of the observed morphology in the visibility domain. We focused on fitting the real part of the visibilities because the emission is mainly axisymmetric. As shown in the bottom right panel of Fig.~\ref{ALMA_SR24S}, the intensity profile along the disk PA is symmetric. When taking the intensity profile along the minor axis of the disk, the difference of emission between the south east and north west is less than 1\,$\sigma$. For all the three observations, the imaginary part of the visibilities  oscillates very close to zero after centering the target (bottom panels of Fig.~\ref{best_fit_models}). 

We considered three different models. First, we assumed a radially asymmetric Gaussian ring for the millimeter intensity with different inner and outer widths. The motivation of this model is to include the effect of particle trapping in a radial pressure bump. \cite{pinilla2017} demonstrated that in the presence of one pressure bump, the accumulation of particles is expected to be radially asymmetric because in the outer disk grains take longer times to grow to sizes for which radial drift is effective. In addition, the drift timescales are longer in the outer disk. As a consequence of these two effects, at million years timescales, the accumulation of dust particles results in a ring with a larger outer width \citep[see also, Fig.~8 in][]{dullemond2018b}. In this case, the intensity profile is given by:

\begin{equation}
I(r)=\begin{array}{rcl}
C\exp\left(-\frac{(r-r_{\rm{peak}})^2}{2\sigma_{\rm{int}}^2}\right) &\mbox{for} & r\leq r_{\rm{peak}}\\
C\exp\left(-\frac{(r-r_{\rm{peak}})^2}{2\sigma_{\rm{ext}}^2}\right) &\mbox{for} & r> r_{\rm{peak}}.
\end{array}
\label{eq:asymmetric_model}
\end{equation}

Our second and third models aim to reproduce the emission from the inner disk as seen in the 2.75\,mm observations. Thus, the second model includes a point source in addition to this radially asymmetric Gaussian; and our third model assumes that the inner emission is a centered Gaussian profile instead of the point source. This inner Gaussian or the point source is multiplied by a factor $A$, which gives the weight of this inner emission with respect to the outer ring.

To fit the data, we used the Markov chain Monte Carlo (MCMC) method, and we used {\it emcee} \citep{foreman2013}. We follow the same procedure as in \cite{pinilla2017}. We explored the free parameters with 200 walkers and 2000 steps in each case. We adopted a set of uniform prior probability distributions for the free parameters explored by the Markov chain, such that

\begin{eqnarray}
r_{\rm{peak}} &\in& [10, 80]\,\rm{au} \nonumber\\
\sigma_{\rm{int}} &\in& [1, 50]\,\rm{au}\nonumber\\
\sigma_{\rm{ext}}&\in& [1, 50]\,\rm{au} \nonumber\\
A &\in& [0, 1]\nonumber \\
\sigma_{\rm{inner disk}} &\in& [0.1, 10]\,\rm{au}\nonumber\\
F_{\rm{total}}&\in& [0.0, 3.0]\,\rm{Jy}
\label{eq:parameter_space}
\end{eqnarray}

We individually performed fits assuming the three different models for each data set. To quantify which of the three models provides a better fit and add a penalty for the number of parameters in the model; we obtain the Bayesian Information Criterion (BIC), which is defined as $\rm{BIC}= \ln(N)N_{\rm{variables}}-2\ln(\hat{L})$, being $N$ the number of data points, $N_{\rm{variables}}$ the number of variable parameters, and $\hat{L}$ the maximum likelihood value. Differences between models of the BIC values between 6 to 10 (or higher) give a strong (or very strong) evidence in favor of the model with the lowest BIC \citep{kass1995}.

The results of this analysis are:

\begin{table*}
\caption{Best model parameters of the MCMC fit}
\label{table:MCMC_fit}
\centering   
\begin{tabular}{|c|c|c|c|c|c|c|c|c|c|}
\hline
\hline       
Band&
$\lambda$&
Lowest BIC&
$r_{\rm{peak}}$&
$\sigma_{\rm{int}}$&
$\sigma_{\rm{ext}}$&
$A$&
$\sigma_{\rm{inner disk}}$&
$F_{\rm{total}}$&
$\sigma_{\rm{ext}}/\sigma_{\rm{int}}$
\\
&
[mm]&
model&
[au]&
[au]&
[au]&
&
[au]&
[mJy]&
\\
\hline
3&2.75 & (C) & 37.10$^{+0.45}_{-0.44}$ & 10.33$^{+0.47}_{-0.47}$ & 13.21 $^{+0.29}_{-0.29}$ & 0.41$^{+0.06}_{-0.08}$ &5.25$^{+0.70}_{-0.57}$ & 30.46$^{+0.15}_{-0.15}$&1.3 \\
\hline
6&1.30 & (C) & 34.49$^{+0.06}_{-0.06}$ & 9.27$^{+0.06}_{-0.06}$ & 15.87 $^{+0.04}_{-0.04}$ & 0.49$^{+0.01}_{-0.02}$&2.75$^{+0.8}_{-0.08}$ & 227.49$^{+0.15}_{-0.15}$&1.7 \\
\hline
9&0.45 & (A) & 30.56$^{+0.32}_{-0.32}$ & 14.27$^{+0.43}_{-0.43}$ & 18.67 $^{+0.19}_{-0.19}$ & ---& ---& 1941.13$^{+4.49}_{-4.49}$&1.3 \\
\hline
\hline
\end{tabular}
\tablecomments{Models: (A) Radially asymmetric Gaussian ring (Eq.~\ref{eq:asymmetric_model}), (B) As Model (A) plus an inner point source, and (C) As Model (A) plus an inner centered Gaussian. The values assumed a distance of 114.4\,pc.} 
\end{table*}

\begin{itemize}
    \item For Band 3 (2.75\,mm), the model with the lowest BIC is the centered inner Gaussian together with a radially asymmetric Gaussian. The differences of the BIC value are $\sim7$ when compared with the model with only the radially asymmetric Gaussian and $>10$ when comparing with the model that assumed the inner disk to be a point source. 
    \item For Band 6 (1.3\,mm), the model with the lowest BIC is also the one that assumed the inner disk to be a centered Gaussian, with BIC differences higher than 10 in both cases. 
    \item For Band 9 (0.45\,mm), the model with the lowest BIC is when only the radially asymmetric Gaussian is assumed. The BIC difference is $\sim$ 5 when compared to the model that includes a point source and $\sim$9 when compared to the model that includes a centered Gaussian. Therefore, in this case the model with the radially asymmetric Gaussian is only modestly preferred. 
\end{itemize}

Figure~\ref{best_fit_models} shows the binned data corresponding to the real part of the visibilities for each wavelength, and  the model with the lowest BIC and the best-fitting parameters, which are summarized in Table~\ref{table:MCMC_fit}. The error bars correspond to the standard error in each bin. In Band 3, the fit of the visibilities recovers a slightly higher total flux than the one obtained from the image directly (30.46\,mJy vs. 28.9\,mJy), which gives a dust disk mass of 58.3$\pm$7.7\,$M_{\oplus}$.

When we checked the residuals (models-observations) in the visibility plane, they are mainly close to zero for the Band 3 and Band 9 observations, but not for the Band\,6 data. In \cite{pinilla2017}, these residuals are attributed to unresolved substructures with the shape of spirals.  The nature of these residuals is discussed in more detail in Sect.~\ref{substructures}.

\begin{figure}
 \centering
   	\includegraphics[width=9cm]{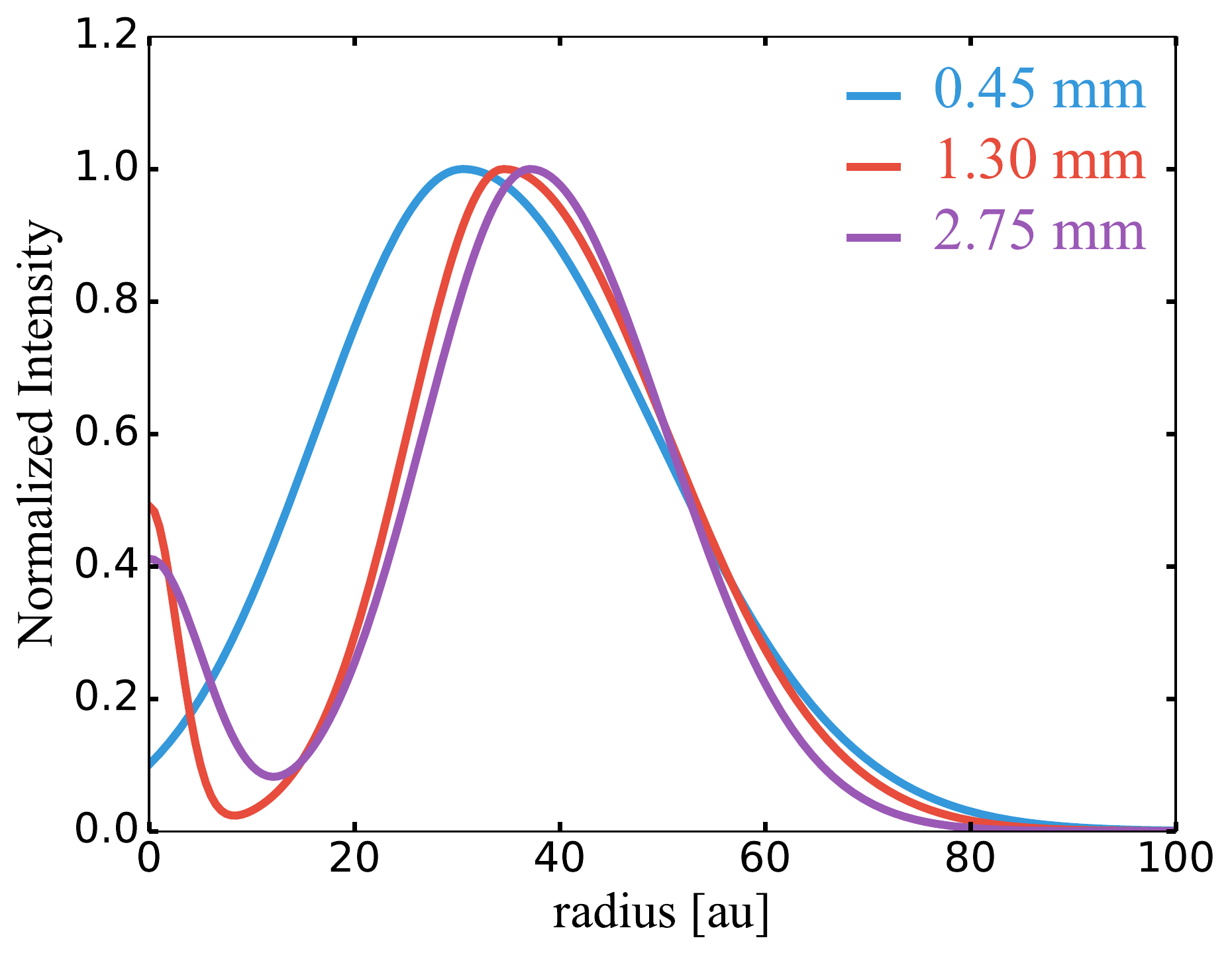}
   \caption{Intensity profiles from models with the lowest BIC and the best fit parameters (Table~\ref{table:MCMC_fit}).}
   \label{intensity_profiles}
\end{figure}

Figure~\ref{intensity_profiles} shows the intensity profile assuming the model with the lowest BIC in each case and the best fit parameters. This fitting analysis shows that the preferred model for the Band\,6 observations includes the inner disk as a centered Gaussian, but this inner disk was not detected in the image. As a test, we used super-uniform weighting, which provides a higher resolution, when cleaning the 1.3\,mm image. However, the inner disk was not detected in the image in this case neither. It is important to note that the inclusion of this inner disk does not help to reduce the residuals obtained in \cite{pinilla2017} at around 800-1000\,k$\lambda$.

The total width of the ring is resolved in Band 6 and Band 3, which is $\sim$25.1\,au in Band 6 (averaged resolution of 19\,au), and  $\sim$23.5\,au in Band 3 averaged resolution of 11\,au). From the results of the fit, the outer ring shows that the internal width of the ring is lower than the external width, i.e., $\sigma_{\rm{int}}<\sigma_{\rm{ext}}$, and that the total width ($\sigma_{\rm{int}}+\sigma_{\rm{ext}}$) decreases at longer wavelength. Both findings are in agreement with particle trapping in a pressure bump. On one hand, in the  outer disk the particles take longer times of evolution to grow to millimeter or centimeter sizes to then drift towards the pressure maximum. As a result, the ring of emission is expected to have an outer tail \citep{pinilla2018}. On the other hand, since larger particles traced at longer wavelengths are drifting more efficiently toward the pressure maximum, the ring-like structure becomes narrower at longer  wavelength. The potential origin of the pressure bump creating this ring is discussed in Sect.~\ref{gap_origin}.

To summarize, in our data, at long wavelength (1.3\,mm and 2.75\,mm), we detect clear evidence for a resolved inner disk up to $\sim$3-5\,au and a radially asymmetric Gaussian ring peaking at $\sim$35-37\,au. At shorter wavelength (0.45\,mm), interestingly, our models do not favor the presence of an inner disk and the ring peaks slightly closer ($\sim$30\,au). 

\section{Discussion} \label{sect:discussion}
\subsection{Origin of the emission from the inner disk} \label{inner_disk}

Centimeter observations of protoplanetary disks have been used to identified ionized jets from weak free-free emission \citep{rodrigues2014, macias2016}, including TDs. \cite{zapata2017} obtained 3.3\,cm observations of 10 TDs with the Jansky Very Large Array (VLA), including SR\,24S to identify potential free-free emission from jets. They compiled data from sub-millimeter to centimeter wavelengths, specifically $\lambda\in[0.088, 0.13, 0.3, 0.73, 0.88, 3.3]\,$cm, to fit the SED with a single or a double power law. With free free emission, the spectral slope of the SED at mm/cm wavelength is expected to become flat. 

For SR\,24S, \cite{zapata2017} found that a two component power law (their Fig.~3) fit the data, with a steeper slope at the sub-millimeter emission, expected from thermal emission from optically thin dust. Specifically, the slope for the centimeter emission (between 0.73 and 3.3\,cm) is 1.46 while for the sub-millimeter emission (between 0.88 and 3\,mm) is 2.89. This value is similar to spatially integrated spectral index calculated in Sect~\ref{dust_spectral}.

We test if we could detect and resolve spatial variations of the spectral index with our current observations. For this, we used the deprojected images that are restored with the same circular beam. However, if there are variations of the spectral index, they remain unresolved in the image plane. As an alternative, we took the best fit models from our visibility analysis in both Band 6 and Band 3 to calculate the total flux in each case within a circle of 20\,au in radius (which encloses mainly the inner disk). 
The spectral index derived from these fluxes is $\sim$2.2, consistent with the value from thermal emission from optically thin dust. From free-free emission the spectral index is expected to be lower than 2. Because of the low value of the spectral index in the inner disk, it is possible that non-dust emission may contribute to this emission. However, since our models favor a resolved inner disk (size of 3-5\,au) instead of an unresolved inner disk in the form of a point source, it is likely that most of this emission is from large grains. This inner disk together with a large gap has also been observed in the TD around T\,Cha \citep{hendler2018}.

The value of the spectral index within the first 20\,au may also indicate that grains have grown to larger sizes (millimeter or centimeter) in the inner disk \citep[e.g.,][]{draine2006} and that they remain there or they are replenished from the outer disk for million years of evolution.  

\begin{figure*}
 \centering
 \setlength{\tabcolsep}{0.1pt}
  \begin{tabular}{cc}   
   	\includegraphics[width=9.0cm]{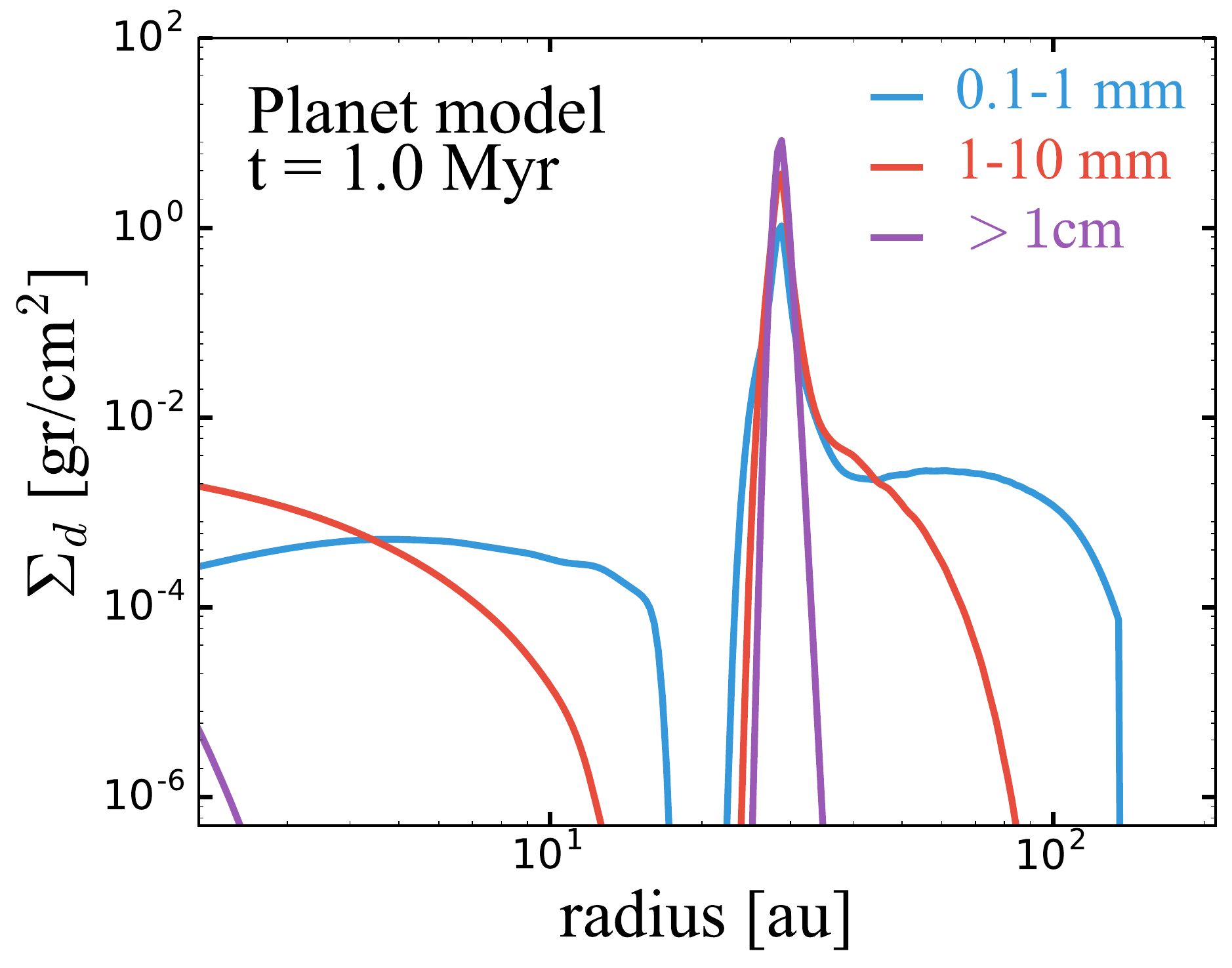}&
	\includegraphics[width=9.0cm]{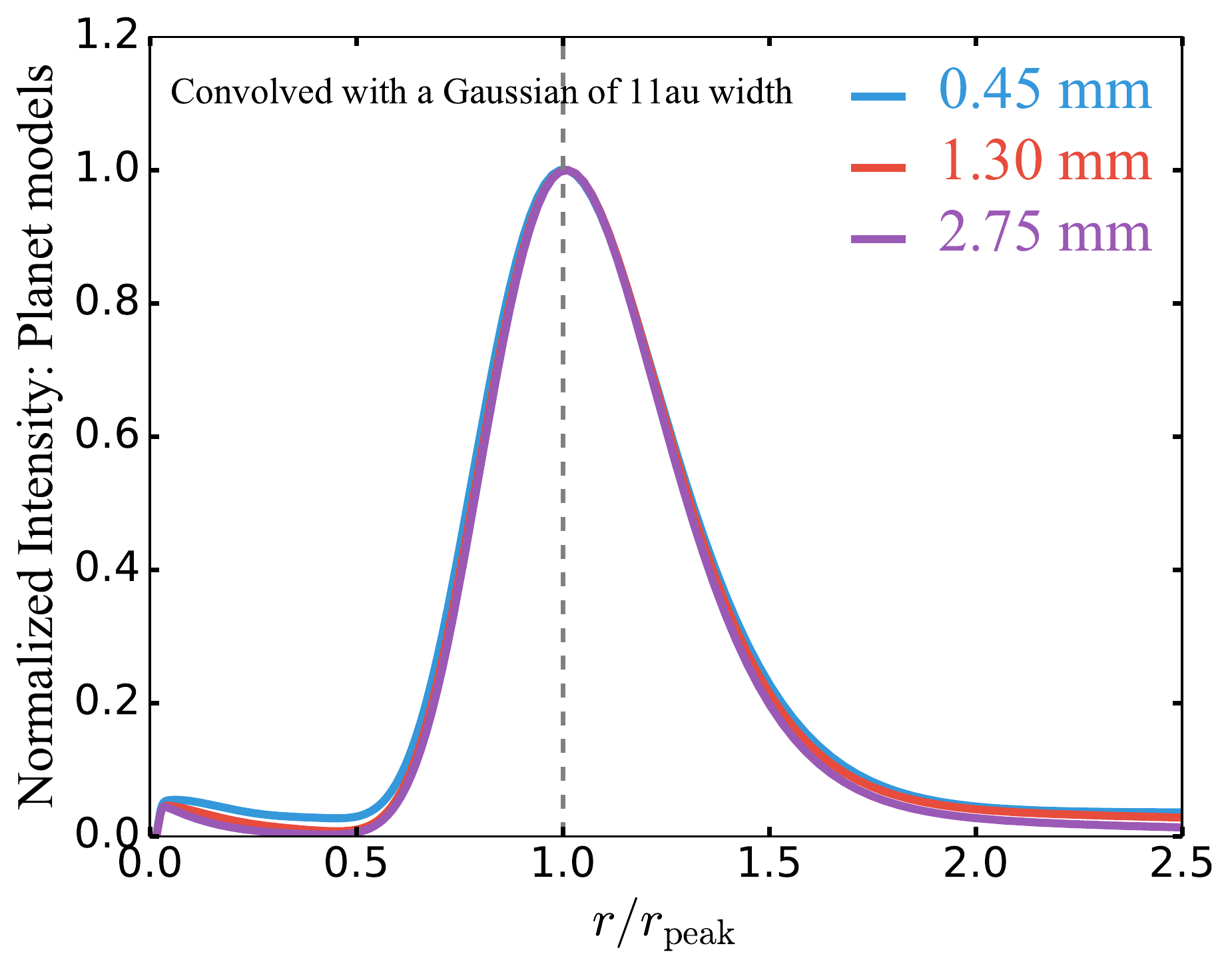}\\
	\includegraphics[width=9.0cm]{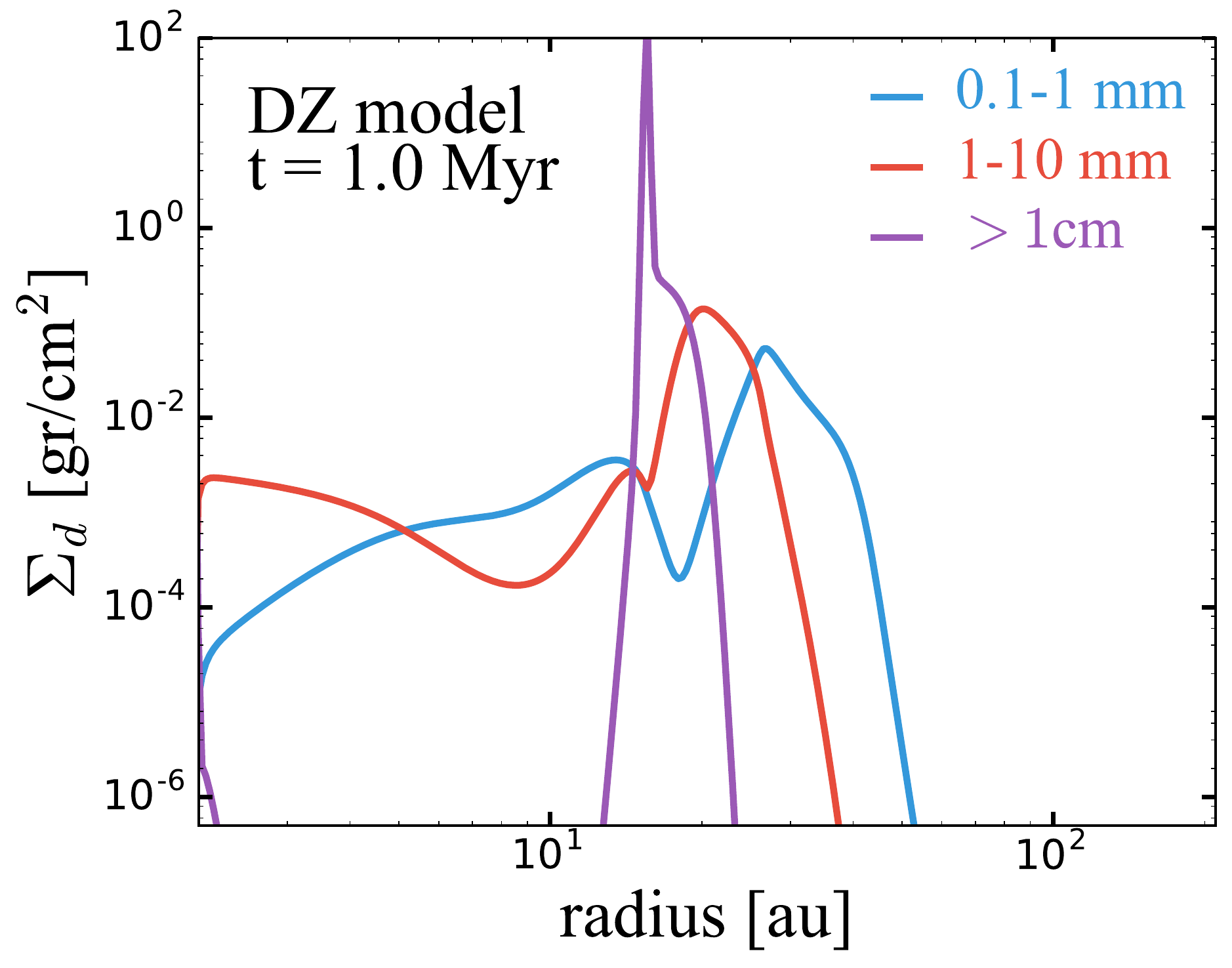}&
	\includegraphics[width=9.0cm]{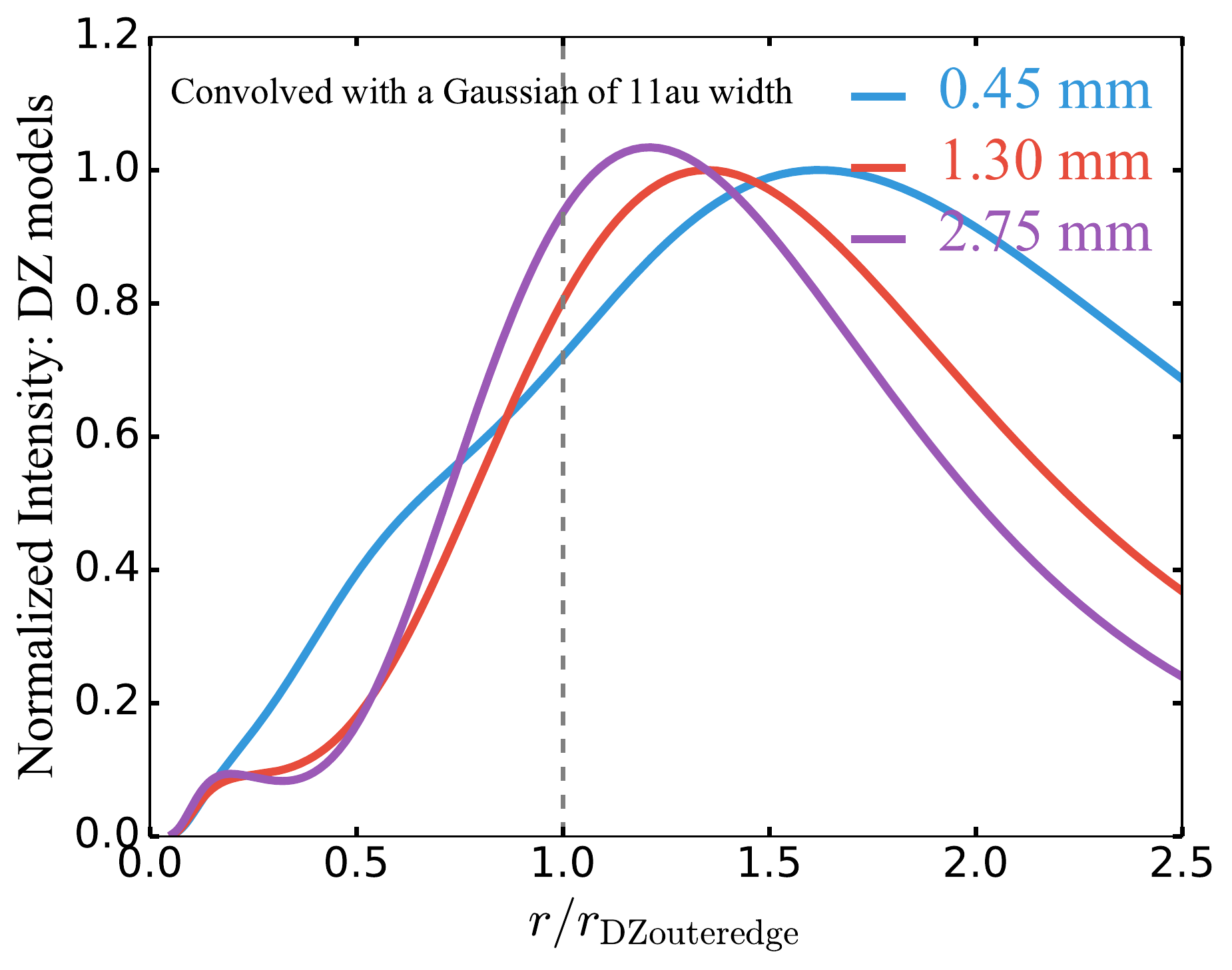}
   \end{tabular}
   \caption{Top panels: dust density distribution for different grain sizes as a function of radius and 1\,Myr of evolution when a 1\,$M_{\rm{Jup}}$ is embedded at 20\,au distance from the star (left), and the corresponding normalized intensity profiles at 0.45\,mm, 1.3\,mm, and 2.75\,mm after convolving with a Gaussian of 11\,au width (right). Bottom panels: as top panels but for the case of a dead zone that is extended up to 20\,au. The details of both models are in \cite{pinilla2012, pinilla2015, pinilla2016a}} 
   \label{model_predictions}
\end{figure*}

\subsection{Origin of the large gap and ring-like structure} \label{gap_origin}

Most protoplanetary disks observed with ALMA at high angular resolution have revealed a variety of sub-structures, being large gaps/cavities and multiple gaps and rings the most common ones \citep[e.g.,][]{long2018}. A large variety of physical mechanism can be responsible for the multiple rings and gaps, including density inhomogeneities (or zonal flows) from the magnetorotational instability, secular gravitational instability, instabilities originating from dust settling, particle growth by condensation near ice lines, planet-disk interaction, among others \citep[e.g.,][]{rice2006, johansen2009,youdin2011, saito2011}. Currently, it is still challenging to observationally distinguish between all these scenarios \citep[e.g.,][]{huang2018}.

However, a few physical mechanisms are currently possible to explain the formation of a large cavity at millimeter emission: the interaction with embedded planet(s) or companion(s) \citep[e.g.,][]{zhu2011}, the existence of a extended dead zone \citep[e.g.,][]{flock2015}, and internal photoevaporation from  stellar irradiation \citep[][]{alexander2007}.

This suggests that disks with a large millimeter-cavity may have a different path of evolution from disks with more millimeter-substructures \citep[multiple rings/gaps, spiral arms, see also][]{garufi2018}, as there are just a few physical processes that lead to large millimeter-cavities.

In the case of internal photoevaporation, models predict a particular combination of cavity size and accretion rate, specifically cavities smaller than around 20\,au with accretion rates lower than $10^{-9}\,M_\odot$year$^{-1}$ \citep[e.g.,][]{ercolano2017}. In the case of SR\,24S, the accretion rate of $\sim3\times10^{-8}$\,$M_{\odot}$\, year$^{-1}$ \citep{natta2006}, and a cavity size of around $\sim$35\,au (Table~\ref{table:MCMC_fit}) exclude the possibility of photoevaporation. Furthermore, photoevaporation predicts a highly depleted cavity in both gas and dust \citep{alexander2007}. Observations of CO and its isotopologues revealed the presence of CO, $^{13}$CO, and C$^{18}$O peaking inside the cavity in the SR\,24S disk \citep{fernandez2017, pinilla2017}, although foreground absorption from the dark cloud in Ophiuchus may affect the distribution of the observed gas emission lines. Nevertheless, our current observations also demonstrate the existence of millimeter sized particles in the inner disk (Sect.~\ref{inner_disk}), which would be difficult to predict in the case of photoevaporation. 

We investigated whether the current ALMA observations of SR\,24S favor one of the two remaining scenarios for cavity formation (dead zone or a massive embedded planet) by qualitatively comparing the dust density distribution predicted by these two models and our current observations. Figure~\ref{model_predictions} show the dust density distribution after 1\,Myr of evolution of different size of dust particles as a function of radius in the case of an 1\,$M_{\rm{Jup}}$ planet embedded a 20\,au distance from the star as compared to the case of a dead zone extended up to 20\,au. The details of these simulations are presented in \cite{pinilla2012, pinilla2015, pinilla2016a}. In short, these models include the transport of the grains (Brownian motion, dust diffusion, settling, and radial/azimuthal drift), as well as the coagulation, fragmentation and erosion of the particles. In the case of an embedded planet in the disk, hydrodynamical simulations are run prior to the dust evolution models until the disk reaches a steady-state for the gas surface density, which is then used as an input for the dust evolution. In the case of a dead zone, a smooth transition in the $\alpha$-viscosity \citep{shakura1973} is assumed at 20\,au. At this location, the change in the gas surface density is such that the disk switch from being active to dead, and back. This transition of $\alpha$-viscosity affects several aspects of the gas and dust evolution, including accretion, turbulence, dust diffusion, dust fragmentation and hence the maximum grain size that particles can reach inside and outside the dead zone. In these two physical scenarios, we expect particle trapping, the formation of a large cavity and a ring like structure at (sub-) millimeter and centimeter wavelengths. In both cases, the expected intensity profiles at 0.45\,mm, 1.3\,mm, and 2.75\,mm are included. For this plot, the intensity profile is normalized to the peak, and the radius is normalized to the location of the pressure maximum ($r_{\rm{peak}}$) or the initial outer edge of the dead zone (20\,au). We convolved these intensity profiles with a Gaussian beam of 11\,au (which is the averaged resolution of the Band 3 observations, 0.95\as, assuming the distance to SR\,24S, i.e., 114.4\,pc), which is the size of the common circular beam used to restore the images with the same resolution at Band 6 and Band 3 (Fig.~\ref{radial_profiles}). To obtain the intensity, we calculated the opacities for each grain at a given wavelength using Mie theory, and assumed optical constants from \cite{ricci2010}. In addition, we took a simple power-law for the radial dependence of the midplane temperature (power-law index of $-1/2$). 

In the planet scenario, a large gap is carved accompanied by a pressure bump at the outer edge, which efficiently traps millimeters/centimeter- sized particles \citep[e.g.,][]{rice2006, gonzalez2012, pinilla2012}.  In this case, the accumulation of large particles (from 0.1\,mm to larger than centimeter) peaks at the pressure maximum and the concentration is narrower for larger particles that are more decoupled from the gas and feel a stronger radial drift toward the pressure maxima.  The degree of radial concentration of large grains relative to small grains is a sensitive function of both planet mass and disk viscosity \citep[turbulently re-mixing the dust, e.g.,][]{dullemond2018b}. Turbulence can affect the gap formation and the concentration of particles such that weak or strong turbulence can lead to a different disk appearance than a cavity and a single ring-liked structure when observed at millimeter-wavelengths \citep{ovelar2016,bae2018}. The intensity profiles at 0.45\,mm, 1.3\,mm and 2.75\,mm are in the planet case very similar after convolving with a Gaussian of 11\,au radial width. In this case the ring-like structure is slightly asymmetric in the radial direction and it has a larger outer width compare to the inner width. By fitting a radially asymmetric Gaussian to these profiles (Eq.~\ref{eq:asymmetric_model}), the ratio of the external to the internal width is $\sim1.4$, similar to the averaged value from our current observations of SR\,24S (Table~\ref{table:MCMC_fit}). The main difference between the predictions of these models and our observations is that in the observations there is a shift of the peak of emission. While the emission at 1.3\,mm and 2.75\,mm peaks almost at the same location (Fig.~\ref{intensity_profiles}), the emission at 0.45\,mm peaks slightly inwards, but this shift is limited by the current data resolution (the minor axis of the beam of the Band 9 observations is 0.19\as, while the shift between Band 9 and Bands 3/6 is around 0.1\as). This shift may result from optically thick emission at 0.45\,mm, which may trace not only variations of the dust distribution, but also of temperature \citep{pinilla2017}.

The predictions for dust cavity formation by a dead zone are different from our observations. In this scenario, the particles grow to larger sizes inside the dead zone where the disk turbulence is lower and the fragmentation of particles decreases. As a  result, the largest particles accumulate closer to the star and the peak of the dust density distribution move inwards for larger grains. This shift would be detectable even at the current resolution of our observations. Note that any pressure bump formed by changes of the disk turbulence could lead to shifts of the peak of the emission at different wavelengths, since the maximum grain size is inversely proportional to the disk turbulence \citep[e.g.,][]{birnstiel2012}. If the bump is formed in a region where the turbulence has a transition from high to low (as the inner edge of a dead zone), the peak is expected to move outwards for longer wavelengths. On the contrary, if the disk turbulence changes from low to high (as at the outer edge of a dead zone) the peak moves inwards for longer wavelengths, as in the case shown in Fig.~\ref{model_predictions}.

\subsection{Limits on the planet mass and disk turbulence}
Our current multi-wavelength observations of SR\,24S favor planet-disk interaction as the main physical mechanism driving the formation of the dust cavity. The observations at  2.75\,mm reveal an inner disk that is likely from dust thermal emission (Sect.~\ref{inner_disk}). This implies that any embedded planet carving the cavity must allow millimeter/centimeter sized particles to remain for million years of evolution in the inner disk. If the planet is very massive ($\gtrsim5\,M_{\rm{Jup}}$), \cite{pinilla2016b} demonstrated that the dust located at the inner disk will drift completely towards the star and that the inner disk will remain empty of dust (of any size) after several million years of evolution ($\sim$5\,Myr). This is because the gap carved by a $5\,M_{\rm{Jup}}$ planet would not allow particles of any size to drift inward, preventing any dust replenishment from the outer to the inner disk. This puts constraints on the upper limit of the mass of any potential embedded planet inside the cavity of SR\,24S.

The value of $5\,M_{\rm{Jup}}$ for the planet mass are for models that assume a disk turbulence of $\alpha\sim10^{-3}$ (assuming $\alpha$ disk viscosity). \cite{ovelar2016} showed that when a massive planet is embedded in the disk, this value of turbulence is needed for the disk to show a cavity and a ring-like structure detectable at millimeter-emission. For a higher disk turbulence, the trapping is not effective and the millimeter emission of dust will be smooth. On the contrary, for low levels of turbulence, the trapping in pressure maxima is so effective that most of the particles grow to very large sizes ($\gtrsim$m) at million years timescales, and these bodies would not emit thermally at millimeter wavelengths. In addition, hydrodynamical simulations showed that the viscous transport in the disk determines the number of gaps that a planet can open \citep[e.g.,][]{dong2017, bae2018}. While a disk viscosity  of $\alpha\sim10^{-4}-10^{-3}$  yields to a single gap, lower viscosities can open multiple gaps and thereby multiple pressure bumps that will create multiple ring like structures at millimeter emission. The current observations of SR\,24S reveal a single ring that favors intermediate values of $\alpha\sim10^{-4}-10^{-3}$, but higher angular resolution observations are required to exclude that this single ring may be a composition of close rings.  

Alternatively, it is possible that the cavity is opened by multiple planets that lead to a shallower gap  in comparison to a single planet with the same mass \citep{duffell2015}. In this case, more dust from the outer disk may tunnel inward, replenishing the inner disk.

\subsection{Unresolved sub-structures} \label{substructures}
Our current observations of SR\,24S show complex residuals when subtracting the ring model from the images, in particular for Band 6 and Band 9 \citep[][see their Fig.~5]{pinilla2017}. We attributed these residuals to unresolved substructures with the shape of spirals. 
It is still possible that these residuals are not seen in the Band 3 observations because the signal to noise ratio is lower for these observations.  In the case of Band 6, the continuum emission is detected with a much higher signal-to- noise ratio with respect to the peak compared to the Band 3 data (256 vs 24). Alternatively, it is possible that the spirals are only detectable when the emission is (partially) optically thick  \citep[as in the case of Band 9 and 6,][]{pinilla2017}, and it is tracing variations of the disk temperature or spiral shocks, potentially from planet-disk interaction \citep{juhasz2015, dong2015, zhu2015}; while larger grains that dominate the 2.75\,mm observations are tracing mainly the ring.

SR\,24S is part of a hierarchical triple system, being SR\,24S the single star. The separation between SR\,24S and the binary system SR\,24N is 5.2\as \citep{reipurth1993}. Recent high angular observations from ALMA reveal spiral arms structures in multiple star systems \citep[e.g.,][]{kurtovic2018}. The same could be happening for the SR\,24 system that shows spiral patterns in scattered light connecting SR\,24S with SR\,24N \citep{mayama2010}. \cite{fernandez2017} found that SR\,24S and SR\,24N disks are strongly misaligned by 108$^\circ$, and they are possibly rotating in opposite directions. This misalignment may explain the origin of the spiral arm connecting the two disks at near infrared emission, although the tidal interaction between disk and star is much weaker if the orbit of the binary and the plane of the disk are misaligned \citep{miranda2015}. These spiral arm structures are not currently seen in the millimeter observations of this system. To determine the nature of these potential substructures in SR\,24S, higher angular resolution and high sensitivity observations at (sub-)millimeter emission are needed.

\section{Conclusions} \label{sect:conclusions}
We report new ALMA Band 3 observations at 2.75\,mm of the TD around SR\,24S with a resolution of 0.106\as$\times$ 0.088\as ($\sim$12$\times$10\,au). We compare our data with previous ALMA observations of the same disk at 0.45\,mm and 1.30\,mm. Our main conclusions are:
\begin{itemize}

\item At 2.75\,mm, we detect a resolved inner disk and a ring-like structure that peaks at $\sim$0.32\as, that is $\sim$37\,au at a distance of 114.4\,pc. The width of this ring is spatially resolved and it is approximately $\sim23$\,au.
    
\item By performing an analysis of the dust morphology at each wavelength in the visibility plane, we found that the total width of the ring like structure decreases at longer wavelength as expected from dust trapping models. In addition, the models favor radially asymmetric rings at the three wavelengths, with larger outer widths (or in other words a ring with an outer tail/wing). These outer wing of the ring is also a natural result of dust trapping since particles take longer times of evolution to grow to millimeter or centimeter sizes in the outer disk to then drift towards the pressure maximum.
    
\item The analysis of the visibilities allow us to conclude that the Band\,6 observations are better representated by a model that included an inner disk. Such inner disk is not currently detected in the image plane from ALMA observations. When calculating the spectral index in the inner disk (inside 20\,au) between the 1.3\,mm and 2.75\,mm, we found that the inner disk emission is likely dominated by dust thermal emission instead of free-free emission. The models do not favor the detection of an inner disk in Band\,9, possibly due to the low resolution in comparison with Band\,3 and Band\,6. Further observations at short ALMA wavelengths  with high angular resolution are needed to test this hypothesis.
    
\item We qualitatively compared the ring morphology of SR\,24S at the three wavelengths with models that predict cavity formation, such as photoevaporation, dead zones, and planet disk interaction. This comparison favors the planet scenario (single or multiple planets).

\item In the case of a single planet inside the cavity of SR\,24S, the existence of an inner disk put constraints on the mass of that potential planet, with an upper limit of $\sim5\,M_{\rm{Jup}}$. The current morphology observed at different wavelength also constrain the disk turbulence, with values of $\alpha\sim10^{-4}-10^{-3}$. Higher or lower values of $\alpha$ would yield to a smooth or multiple rings/gaps distributions that are not yet seen in this disk, respectively.
\vspace{-0.2cm}
\item Future higher angular resolution and high sensitivity observations at (sub-)millimeter emission are needed to investigate the existence of potential spiral arms in SR\,24S, potentially originated by the multiplicity of the system. Currently, such structures are not observed and only hints remain in the analysis of the visibilities of the 1.3\,mm data. 
\end{itemize}

\software{CASA \citep{mcmullin2007}, emcee \citep{foreman2013}}

\paragraph{Acknowledgments}
  \acknowledgments{We thank the referee R. Dong for his prompt and constructive referee report. P.P. acknowledges support by NASA through Hubble Fellowship grant HST-HF2-51380.001-A awarded by the Space Telescope Science Institute, which is operated by the Association of Universities for Research in Astronomy, Inc., for NASA, under contract NAS 5-26555. D.H. is supported by European Union A-ERC grant 291141 CHEMPLAN, NWO and by a KNAW professor prize awarded to E. van Dishoeck. D.H. is part of Allegro, the European ALMA Regional Centre node in the Netherlands funded by the Netherlands Organisation for Scientific Research (NWO). L.M.P. acknowledges support from CONICYT project Basal AFB-170002 and from FONDECYT Iniciaci\'on project \#11181068. This paper makes use of the following ALMA data: ADS/JAO.ALMA\#2017.1.00884.S. ALMA is a partnership of ESO (representing its member states), NSF (USA) and NINS (Japan), together with NRC (Canada) and NSC and ASIAA (Taiwan) and KASI (Republic of Korea), in cooperation with the Republic of Chile. The Joint ALMA Observatory is operated by ESO, AUI/NRAO and NAOJ.} 
  
\appendix
\section{Band 3 Residuals}
\begin{figure}
 \centering
   	\includegraphics[width=9cm]{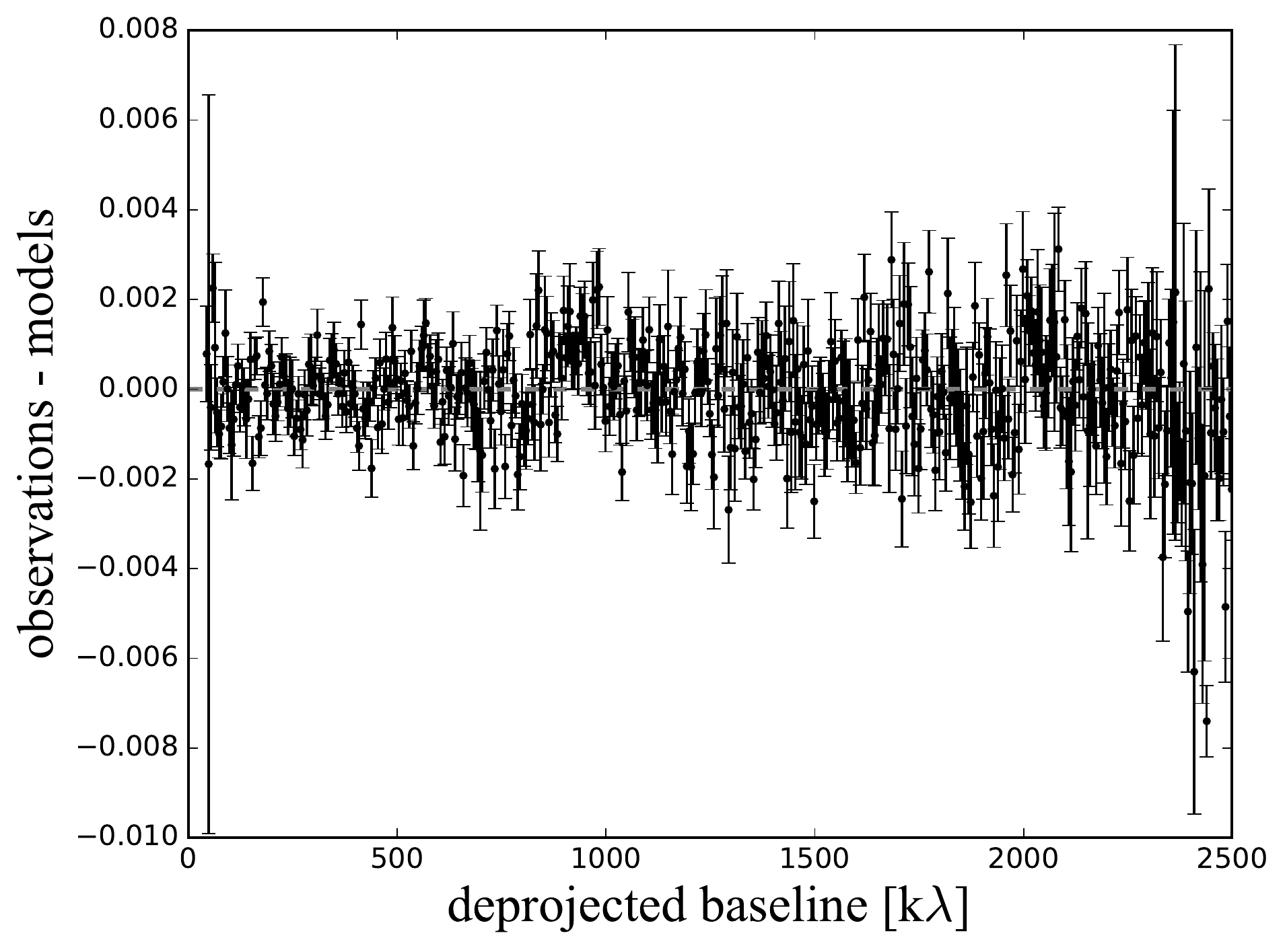}
   \caption{Band 3 residuals (observations-models) when taking the best model fit shown in Fig.~\ref{best_fit_models}. }
   \label{residuals_B3}
\end{figure}

Figure~\ref{residuals_B3} shows the Band 3 residuals (observations-models) when taking the best model fit shown in Fig.~\ref{best_fit_models}, i.e. the model with a radially asymmetric Gaussian and an inner centered Gaussian. The residuals for Band 6 and Band 9 were shown in \cite{pinilla2017} and remain similar for both Bands despite the inclusion of an inner Gaussian for the model in Band 6. 

\bibliographystyle{aasjournal}


\end{document}